# Real-time deep learning phase imaging flow cytometer reveals blood cell aggregate biomarkers for haematology diagnostics

Kerem Delikoyun[1,2,*], Qianyu Chen[1,2], Liu Wei[3], Si Ko Myo[2], Johannes Krell[4], Martin Schlegel[4], Win Sen Kuan[5,6], John Tshon Yit Soong[5,7], Gerhard Schneider[4], Clarissa Prazeres da Costa[8,9,10], Percy A. Knolle[11], Laurent Renia[12,13,14], Matthew Edward Cove[7], Hwee Kuan Lee[3,15,16,17,18], Klaus Diepold[1,2], Oliver Hayden[1,2,*]

[1] School of Computation, Information and Technology & Munich Institute of Biomedical Engineering, Technical University of Munich, TranslaTUM, Einsteinstraße 25, 81675 Munich, Germany
[2] TUMCREATE, 1 Create Way, 138602, Singapore
[3] Bioinformatics Institute, Agency for Science, Technology and Research, 30 Biopolis Street, 138671, Singapore
[4] Department of Anesthesiology and Intensive Care Medicine, TUM University Hospital, Munich, Germany
[5] Yong Loo Lin School of Medicine, National University of Singapore, 10 Medical Dr, 117597, Singapore
[6] Emergency Medicine Department, National University Hospital, 5 Lower Kent Ridge Road, 119074, Singapore
[7] Department of Medicine, National University Hospital, NUHS Tower Block, 119228, Singapore
[8] Center for Global Health, School of Medicine and Health, Technical University of Munich, Munich, Germany
[9] Institute for Medical Microbiology, Immunology and Hygiene, School of Medicine and Health, Technical University of Munich, Munich, Germany
[10] German Center for Infection Research (DZIF), Partner Site Munich, Munich, Germany
[11] Institute of Molecular Immunology, School of Medicine and Health, Technical University of Munich, Munich, Germany
[12] Infectious Diseases Labs (AIDL), Agency for Science, Technology and Research (A*STAR), Singapore
[13] Lee Kong Chian School of Medicine, Nanyang Technological University, Singapore
[14] School of Biological Sciences, Nanyang Technological University, Singapore
[15] Centre for Frontier AI Research, Agency for Science, Technology and Research, 1 Fusionopolis Way, 138671, Singapore
[16] International Research Laboratory on Artificial Intelligence, Agency for Science, Technology and Research, 1 Fusionopolis Way, 138671, Singapore
[17] School of Biological Sciences, Nanyang Technological University, 60 Nanyang Dr, 639798, Singapore
[18] School of Computing, National University of Singapore, 13 Computing Dr, 117417, Singapore

*Corresponding authors: kerem.delikoyun@tum-create.edu.sg, oliver.hayden@tum.de

**Abstract**

While analysing rare blood cell aggregates remains challenging in automated haematology, they could markedly advance label-free functional diagnostics. Conventional flow cytometers efficiently perform cell counting with leukocyte differentials but fail to identify aggregates with flagged results, requiring manual reviews. Quantitative phase imaging flow cytometry captures detailed aggregate morphologies, but clinical use is hampered by massive data storage and offline processing. Incorporating "hidden" biomarkers into routine haematology panels would significantly improve diagnostics without flagged results. We present RT-HAD, an end-to-end deep learning-based image and data processing framework for off-axis digital holographic microscopy (DHM), which combines physics-consistent holographic reconstruction and detection, representing each blood cell in a graph to recognize aggregates. RT-HAD processes >30 GB of image data on-the-fly with turnaround time of <1.5 min and error rate of 8.9% in platelet aggregate detection, which matches acceptable laboratory error rates of haematology biomarkers and solves the "big data" challenge for point-of-care diagnostics.

# Introduction

Sepsis and thrombo-inflammatory disorders remain leading causes of morbidity and mortality, in part due to the formation of blood cell aggregates such as platelet-platelet (PP), leukocyte-leukocyte (LL), and leukocyte-platelet (LP) interactions, which can be linked to disease severity and serve as promising early biomarkers for point-of-care testing (POCT) in the acute care [1-3]. In sepsis, dysregulated immune responses trigger immunothrombosis, and similar aggregate patterns have been observed in COVID-19 patients in the ICU with septic pulmonary complications [4]. However, these aggregates remain undetectable by standard automated hematology analyzers [5-7]. Haematology analysis, the most frequently requested in vitro diagnostic test, is a cornerstone of clinical diagnostics and provides complete blood counting (CBC) and differential leukocyte counts (Diff). While traditional automated analyzers like Coulter counters and scatter-based flow cytometers offer a turnaround time (TAT) of ~30 sec [8], they offer limited morphological information, cannot resolve cellular aggregates, and lack functional diagnostic capacity [9]. Consequently, diagnostically unique insights such as PP microaggregates remain hidden in routine assessments [7, 10].

Blood cell subtyping with fluorescence flow cytometry and specific antibodies is currently the only opportunity for high-throughput functional cell analysis. Since traditional flow cytometers struggle when it comes to the identification of blood cell aggregates [11], imaging flow cytometry (IFC), such as Cytek's Amnis Imagestream [12], bridges this gap by combining flow cytometry's throughput with high-resolution fluorescence imaging [13]. IFC captures spatial and morphological features at single-cell resolution [14], enabling aggregate detection. However, its clinical application is hindered by intensive sample preparation and data acquisition (>60 minutes), manual gating and data analysis (30-60 minutes) [15]. High throughput is essential for detecting rare cell populations but poses significant challenges in data storage, transfer, and analysis [16, 17]. Despite fluorescent labelling enabling detailed cell subtyping, IFC remains a research tool, not applicable for clinical use due to limitations in TAT, standardization, automation, and cost [18].

Quantitative phase imaging (QPI) has emerged to overcome IFC's clinical limitations with a high-throughput and label-free cellular analysis workflow, requiring no sample preparation [19-21]. Among the various QPI techniques, digital holographic microscopy (DHM) stands out for clinical settings due to its robustness by encoding phase into an intensity-only hologram, enabling computational reconstruction of both amplitude and phase at high contrast not achievable by bright-field microscopy [22-24]. This single-shot imaging approach yields rich morphological cellular data (e.g., shape, granularity, and intracellular variations) [25, 26]. Importantly, DHM can be implemented in high-speed flow systems with automated microscopy, and its use of viscoelastic flow in microfluidic chips allows parallel imaging without the high-shear serial analysis of conventional hematology analyzers [27], thus preserving fragile blood cell aggregates [4, 28]. Early studies have demonstrated DHM's potential in hematology for label-free leukocyte differentiation and even leukemia subtyping from phase signatures [4, 19, 29].

Traditional image processing methods fall short in high-throughput imaging, which demands frameworks optimized explicitly for fast image and data analysis. Deep learning-based computer vision methods have significantly improved the speed and accuracy of cellular analysis in biomedical imaging [30-33]. Convolutional neural networks (CNNs) have proven effective for cell detection and localization in microscopic images [34]. However, region proposal-based models using Mask R-CNN remain computationally burdensome due to their heavy backbones and multi-stage proposal generation, up to 4× slower than recent architectures such as YOLOv8 [35]. Yet even these fast and efficient algorithms alone are insufficient to develop clinically relevant diagnostic applications due to DHM's computationally intensive holographic reconstruction process before object recognition [36-38]. Conventional preprocessing steps such as thresholding, feature extraction, and ROI-based patch generation further hinder DHM's potential in clinical high-throughput imaging, due to their time-consuming and complex nature [4].

Another bottleneck for applying DHM in haematology is data storage. The high acquisition rate of DHM (105 frames/second) typically generates more than 30 GB of raw data/patient, which prevents its routine use in real-time haematology analysis for clinical settings [20, 39]. For a real-world clinical application, an automated system for a CBC/Diff with functional blood cell aggregates will require petabyte scale storage, assuming that 3,000 CBCs/day performed in a large central laboratory generating data equivalent of 30 PB/year, which poses a significant "big data" problem [40, 41], illustrated in Figure 1. Without a real-time analysis option, the storage requirements for batch processing would become excessively costly and increase the carbon footprint of DHM as a medical imaging tool [42]. Unlike other medical imaging methods, such as MRI or CT scans, which typically must be stored for over five years in lossless formats like DICOM (subject to varying country-specific regulations) to facilitate future clinical reviews, the raw machine-generated data produced in haematology analyses does not fall under these storage mandates. This regulatory leniency makes haematology particularly well-suited for implementing an AI-powered image and data processing framework that offers real-time analysis without any raw data storage requirement [43, 44].

We introduce RT-HAD: Real-Time Holographic Aggregate Detector, an end-to-end image and data processing framework that eliminates key barriers to deploying DHM in clinical haematology. RT-HAD combines multiple specialized deep learning models to achieve real-time, quantitative analysis of single blood cells and blood cell aggregates, revealing hidden haematology biomarkers with high clinical utility. RT-HAD integrates three specialized deep learning modules: *i)* a model for holographic reconstruction and phase retrieval, *ii)* an object detection model for individual blood cell identification, and *iii)* a graph-based aggregate analyzer to detect blood cell aggregates by representing each cell as a node. The framework processes each raw hologram in under 10 ms, introducing only minimal latency relative to the DHM acquisition rate (9.5 ms/frame), which is mitigated by buffered transfer. RT-HAD achieves an error rate of only 8.9% in platelet aggregate detection compared to human experts, enabling results parallel with sample acquisition (~1.5 min). Furthermore, the framework is microscopy agnostic and can be applied to different QPI platforms with little to no modifications to accelerate translational research.

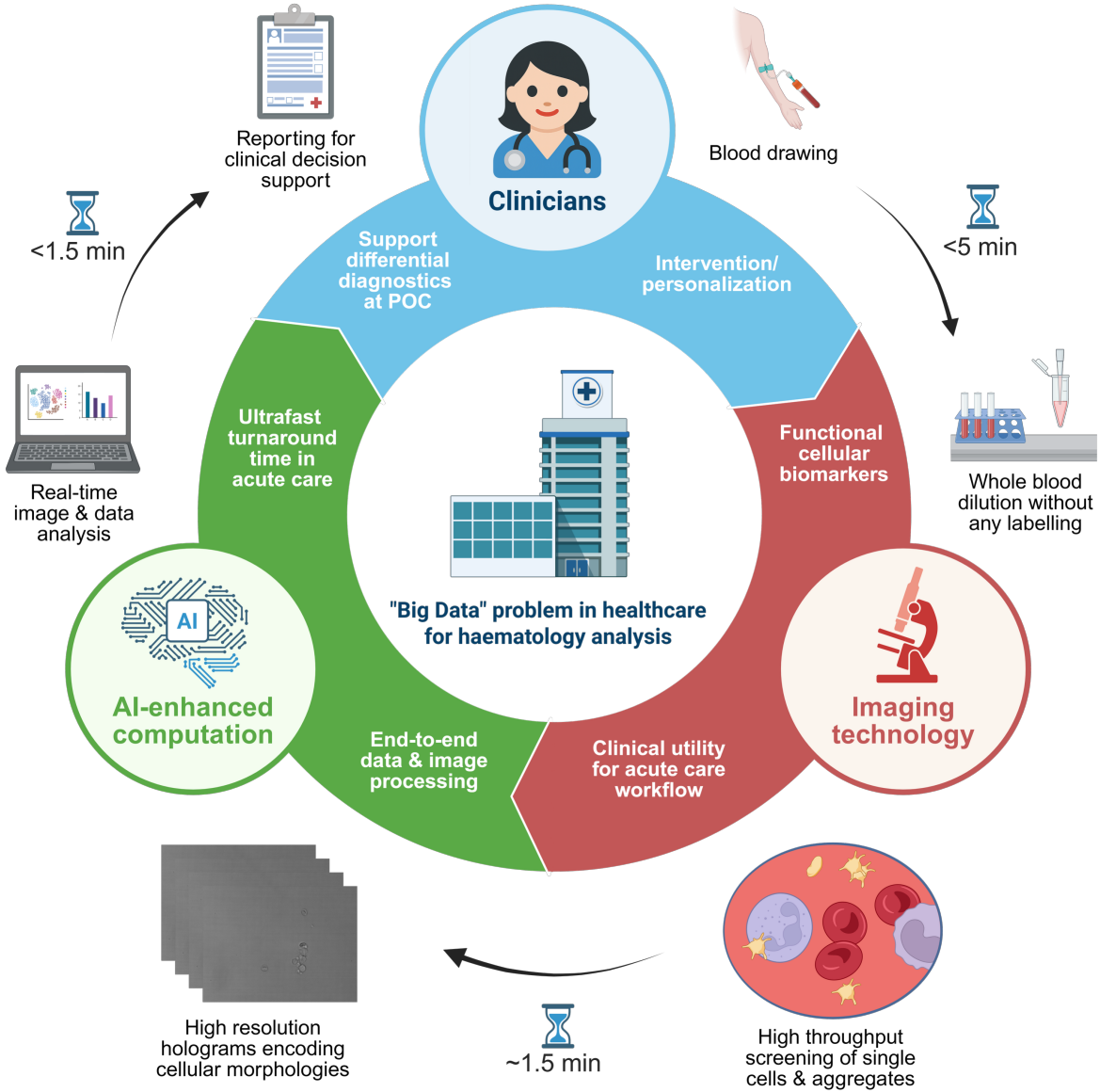

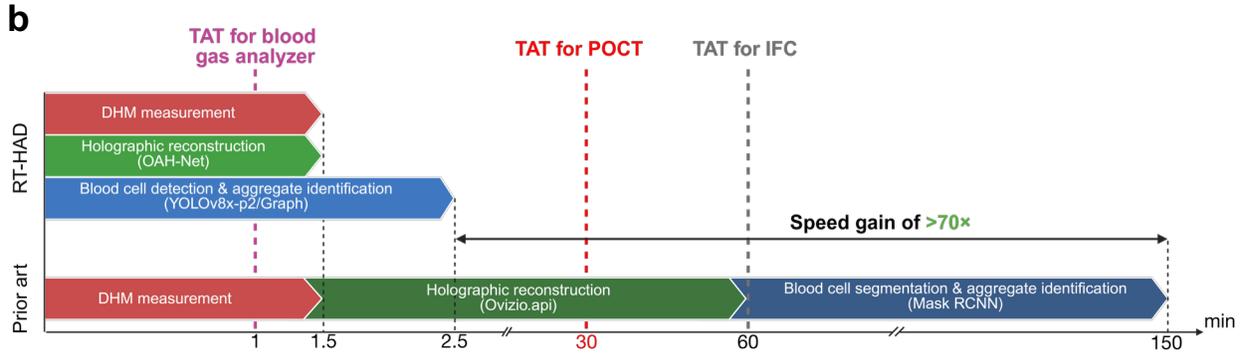

**Figure 1: Clinical workflow for real-time aggregate detection. (a)** Overview of the big data challenge posed by modern medical imaging technologies in healthcare, emphasizing the need for efficient and accurate differential diagnostics at the point-of-care for personalized interventions. DHM provides label-free imaging, capturing functional cellular biomarkers useful as early predictive indicators in acute care settings. RT-HAD system addresses these challenges by integrating advanced deep learning algorithms for rapid, real-time data analysis, significantly reducing turnaround time (TAT). This is particularly advantageous in haematology, where regulatory guidelines do not mandate long-term storage of raw imaging data, thereby promoting AI-driven imaging and data processing solutions. **(b)** Comparative analysis illustrating the advantages of the RT-HAD system over other medical technologies: 30 min of TAT for POCT is standard. At one end, blood gas analyzers, one of the most common POC tests in clinical settings, have a TAT of 1 min, and on the other end, IFC has a TAT of >60 min. RT-HAD leverages AI-driven holographic reconstruction and advanced blood cell aggregate identification algorithms, enabling the quantification and analysis of PP aggregates in acute-care patients with an ultrafast TAT (<1.5 min). Prior art for DHM relying on standard image-processing techniques typically requires significantly longer processing times (~2.5 h) [4], highlighting RT-HAD's speed gain of more than 70-fold.

## Results

To enable real-time, high-throughput haematology analysis with DHM-based imaging flow cytometers, we developed RT-HAD, an end-to-end framework optimized for speed and clinical relevance. The architecture integrates three key components: holographic phase reconstruction, a deep learning-based object detector for single-cell recognition, and a graph-based aggregate analyzer for identifying clinically significant blood cell aggregates (Fig. 2). RT-HAD is specifically designed to minimize inference time through efficient data handling and targeted analysis of PP aggregates.

### RT-HAD's architectural components and performance assessments

In the first stage, we utilize OAH-Net [39], a physics-consistent holographic reconstruction model that we developed in our previous study, in which raw holograms are rapidly converted into high-resolution phase and amplitude images with integrated phase retrieval in the first stage. OAH-Net is composed of two distinct neural network modules. The first module spatially filters the $x$ and $y$ components of the object wave in Fourier space using learnable filter matrices, thereby isolating the interference term. The second module scales and converts the resulting complex-valued output into separate phase and amplitude images. This integrated process yields a mean absolute error (MAE) across the entire frame of 0.012 for phase images and 0.372 ± 0.016 for amplitude images between ground truth and OAH-Net, with a mean inference time of 4.7 ms/frame over 10,000 frames. This design speeds up the reconstruction process and enables high detection accuracy by providing information-rich phase images that reveal the morphological features of blood cells.

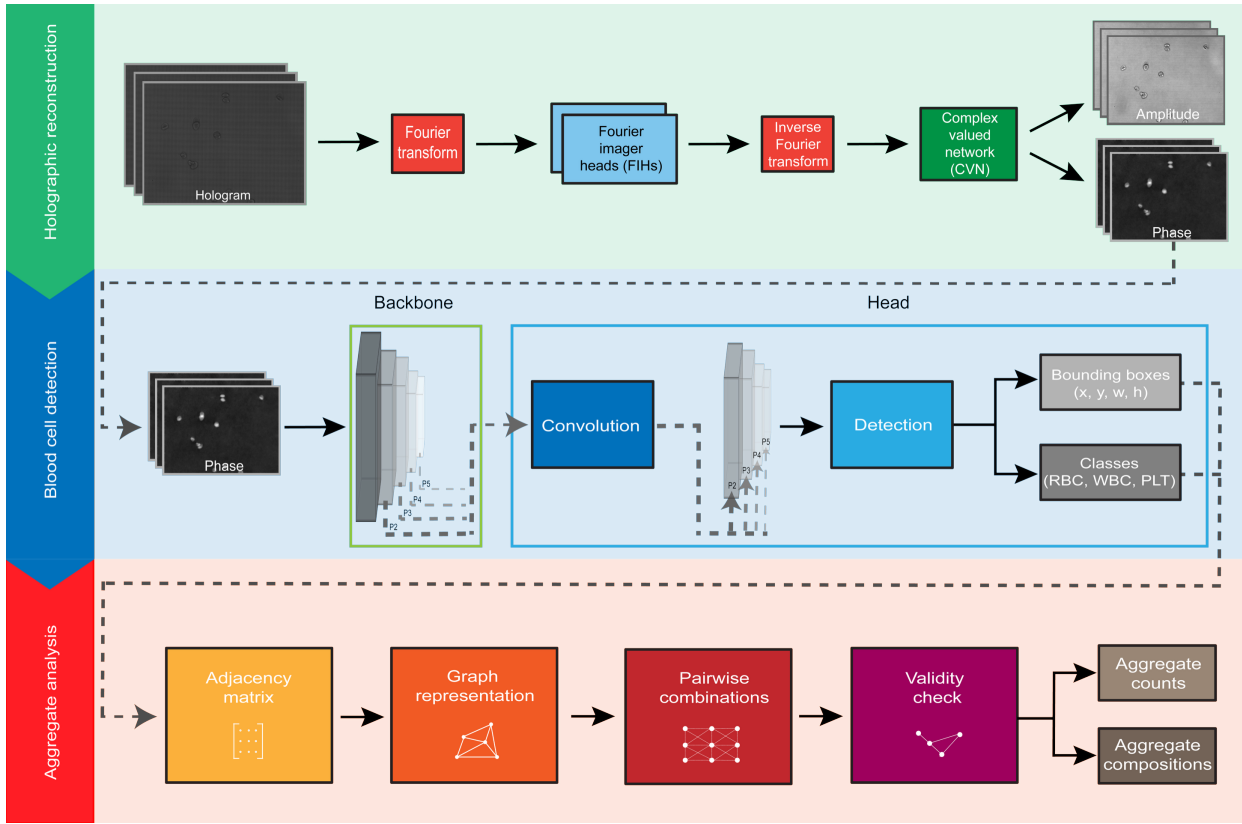

**Figure 2: Detailed architecture of the RT-HAD framework.** Schematic diagram illustrating the integrated RT-HAD framework, composed of three core components: *i)* holographic reconstruction: OAH-Net [39] rapidly reconstructs raw holograms into amplitude and phase images, *ii)* blood cell detection: YOLOv8x-p2-based detection module identifies and classifies individual blood cells, significantly enhancing accuracy for small-sized cells such as platelets, alongside erythrocytes and leukocytes for precise localization with high-throughput imaging analysis and *iii)* aggregate analysis: A graph-based analytical framework translates detected blood cell positions into an adjacency matrix, representing spatial relationships to identify valid aggregates. This integrated AI-driven architecture achieves sub-10 ms inference time while dramatically reducing data storage demands (>99%), enabling on-the-fly analysis with an enhanced haematology biomarker panel for the POC.

In the second stage, RT-HAD performs object detection in 6.6 ms per frame without any quantization strategy. However, an important performance metric for image-based haematology analysis is the accurate recognition of small platelets. In our experiments, RT-HAD achieved both precision and mAP50 of 96.8% in detecting platelets on the holdout test set. We further compared the accuracy of blood cell detection and quantification between DHM and a commercial automated haematology analyzer (Sysmex XN-350, Japan). The ratios among detected blood cells at DHM were found to be very consistent with an automated haematology analyzer, with differences in ratios in erythrocytes and platelets being less than 1%. In comparison, it was less than 0.2% for leukocytes (Fig. S1). Therefore, RT-HAD's detection and quantification accuracy for different blood cell types closely matches a standard automated haematology analyzer used in clinical routines.

In the third stage, after recognition of single cells in the phase images, RT-HAD further identifies cells forming aggregates and determines the aggregate types, such as PP or LL (Fig. 3). To quantify the accuracy of aggregate identification, we compared the system's predicted PP aggregate counts against ground truth provided by human experts directly on a clinical patient sample with error rate of only 8.9%, meaning the majority of true PP aggregates were correctly identified. The misclassification rate (<10%, compared to human experts) can be attributed to biological variance and is small compared to the level of PP aggregates that patients struggle with severe infections reported in the literature, with over 50% of all platelets [4]. The correspondence between predicted and accurate PP aggregate counts is illustrated in Figure S2.

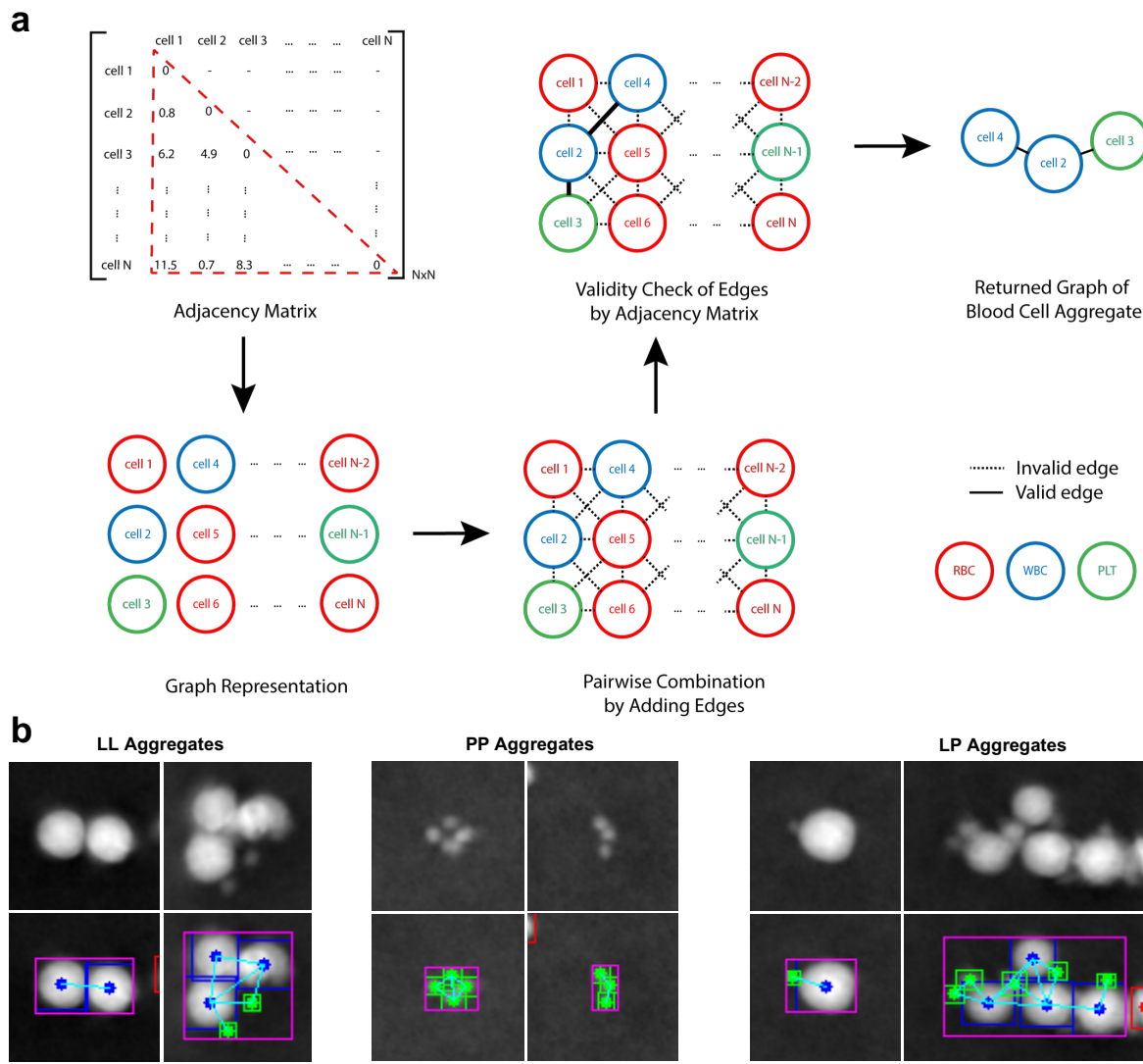

**Figure 3: Aggregate analysis and quantitative visualization. (a)** Graph representation of blood cells, in-depth view of the graph model used for aggregate analysis, illustrating how the adjacency matrix is generated to represent each cell as a node in the graph, and valid edges between pairwise combinations to detect LL, PP, and LP aggregates even though the structure might contain erythrocytes. **(b)** High-resolution exemplary phase images of cell aggregates of different types of LL, PP, and LP aggregates are on top, and the graph model in the bottom images identifies aggregates. The size bar represents 10 μm.

In our framework, the hologram reconstruction is a significant step for effective downstream blood cell detection and aggregate analysis tasks. Although raw holograms of 1,536 × 2,048 pixels with 4-fold oversampling contain all the optical information encoded, the interference fringe patterns obscure small features such as platelets, making the direct use of holograms for object detection both computationally intensive and dramatically less accurate. In our experiments, performing object detection directly on raw holograms was found to be dramatically slower (80.8 ± 0.1 ms/frame) compared to on phase images due to preprocessing steps, while delivering significantly worse performance in the detection of platelets with precision and mAP50 of 77.7% and 71.4%, respectively (Table 1).

**Table 1 | Model performance and inference times for different models and modalities on test set given in mean ± SD.**

| Model | Params (M) | Image modality | Image type | Size (pixels) | Precision -all (%) | mAP50 -all (%) | Speed (ms) |
|---|---|---|---|---|---|---|---|
| Mask R-CNN | 44.0 | Phase | Patch | 96 × 96 | 82.8 ± 2.5 | 99.0 ± 0.5 | 34.9 ± 0.2 |
| YOLOv8x-p2 | 66.6 | Hologram | Full | 1,536 × 2,048 | 79.6 ± 2.1 | 88.9 ± 0.9 | 80.8 ± 0.1 |
| YOLOv8x-p2 | 66.6 | Phase | Full | 384 × 512 | 93.8 ± 1.4 | 96.3 ± 0.6 | 6.6 ± 0.2 |

**Ablation study against the prior art framework**

To evaluate the effectiveness of RT-HAD's object detection performance, we compared its approach of processing full-size reconstructed phase images in a single pass without any preprocessing against the prior framework that uses conventional holographic reconstruction (e.g., angular spectrum method (ASM)) followed by patch extraction and individual cell classification [35]. In the prior art, holographic reconstruction and phase unwrapping take >300 ms/frame, followed by the generation of patch images containing single blood cells or aggregates for object recognition. More precisely, the inference time in prior art for single-cell and aggregate identification takes 34.9 ± 0.2 ms per patch (96 × 96 pixels). A full-size phase image (384 × 512 pixels) contains an average of 25 cells. Therefore, if each cell were processed as its patch, the inference time for a single image could exceed 1 s. It indicates that RT-HAD enables speed gain of more than 5× only in object detection per inference run, and more than 70-fold for the entire image processing framework while increasing the throughput by 25-fold per inference. Regarding platelet detection performance, RT-HAD achieved a precision and mAP50 of 96.8%; in contrast, the prior art framework showed a substantial drop in precision, reaching only 63.9%, despite maintaining a comparable mAP50 of 98.2%. These results confirm that using reconstructed full-size phase images accelerates the inference process and substantially improves the detection accuracy for small objects (e.g., platelets), compared to patch-based analysis relying on traditional object detection models (e.g., Mask R-CNN).

We benchmarked the processing speed of each component of RT-HAD to assess its alignment with TAT for clinical standards in POCT (<30 min) [45]. Our in-house developed holographic reconstruction and phase retrieval algorithm, OAH-Net [39], leverages physics-consistent holographic reconstruction to generate phase and amplitude images in 4.7 ms/frame. The detection model then processes the reconstructed phase images to locate cells and determine their types. Since we are mainly interested in PP aggregates, the detection model was chosen and specifically optimized for high accuracy in detecting small objects. Despite the whole frame and numerous objects per frame, the model maintains this low latency to reach 6.6 ms/frame in complete precision (i.e., FP32). We further achieved faster inference by quantizing the detection model weights to run inference on half-precision (i.e., FP16), which yielded 4.7 ms/frame. Consequently, the quantization of YOLOv8x-p2 resulted in an average inference speed gain of over 25%. Therefore, we deployed the deepest YOLOv8x-p2 architecture due to significant performance improvements on platelet detection despite a slight increase in inference time (Table S2). However, full-size phase images are processed in under 10 ms, fast enough to match the DHM imager's 105 FPS acquisition rate, with data transfer latency effectively buffered to maintain seamless performance.

Thereafter, the recognized individual blood cells are analyzed to identify the blood cell aggregates by checking the validity of the spatial relations in the graph. This module represents each cell as a node in the graph, groups the detected cells based on spatial proximity, and checks their class labels to categorize each cell-cell pairing to include only target classes (i.e., leukocytes and platelets). The aggregate analysis module is very lightweight and integrated into the detection model. In case of extremely high concentration of target cells per frame, the inference time could increase slightly due to graph complexity and reach around 1 ms/frame. Still, on average, over 10,000 frames/measurement in a typical clinical patient measurement, the aggregate identification speed remains at 0.5 ms/frame. Table 2 compares the traditional data processing approach and RT-HAD's real-time and on-the-fly analysis capability, eliminating raw data storage.

**Table 2 | Data processing speed comparison between RT-HAD and prior art [4] in holographic reconstruction, blood cell detection, and aggregate identification with storage requirements.**

| Framework | Inference speed (per frame) | | | | | Storage | | |
| --- | --- | --- | --- | --- | --- | --- | --- | --- |
| | Recon (ms) | Detect (ms) | Agg (ms) | Total (ms) | Real time | Data format | Size (MB) | Storage |
| Prior art | 300 | ~900 | - | >1,000 | No | h5 | ~30,720 | Raw data |
| RT-HAD | 4.7 | 4.7 | 0.5 | <10 | Yes | Tensor | ~15 | Analysis results |

Moreover, a key advantage of RT-HAD's workflow is the substantial reduction in data storage requirements achieved by storing only the detected cell regions for target cells (e.g., platelets and leukocytes) in phase images instead of entire holographic frames. We quantified the storage savings of this region-of-interest (ROI) storage strategy in a typical use case. In our experiments, a raw measurement file containing stacks of full-size holographic frames (i.e., 1,536 × 2,048 - 16 bit), corresponding to 10,000 frames per measurement, is roughly 30 GB. By contrast, RT-HAD processes each frame on-the-fly and saves only the relevant ROIs containing target blood cells and aggregates (i.e., leukocytes and platelets), each cropped around the cells/aggregates with a size of 96 × 96 pixels, and all ROIs are stored in a single h5 file with analysis results and metadata. On average, a typical measurement of a healthy individual yields 3,000 platelets, 200 leukocytes, and fewer than 150 aggregates of all types in total. Storing all ROIs turns into a single h5 file with a size of only ~15 MB. This results in an over 99% reduction in data storage size per measurement compared to storing raw measurements containing holograms for offline processing. This immediately turns out to be a substantial economic and engineering benefit, not to mention any mass storage hardware accompanying DHM. Still, if necessary, only clinically relevant cell samples are stored for further examinations instead of all the analyzed measurements. The storage efficiency directly benefits clinical deployment with a reduced carbon footprint for DHM as an eco-friendly and sustainable medical imaging technology.

**Validation of RT-HAD's clinical utility on quantification of platelet microaggregates as early predictive biomarkers in an acute-care setting**

We further investigated RT-HAD's clinical utility in quantifying platelet aggregate concentration and composition as a potential prognostic biomarker for risk stratification in acute care. We compared the results with the sequential organ failure assessment (SOFA) score. Using our DHM measurements analyzed by RT-HAD, we longitudinally observed the microaggregate formation in a pneumonia cohort from the ICU. The baseline platelet aggregate levels for a healthy patient cohort as control show mean platelet aggregate ratios of 2.0 ± 1.1% (n=10) with >82.7% of microaggregates having two platelets in size. We define an aggregate from platelets as a minimum of two interacting cells. In previous work, we have shown that with a 1:100 dilution of whole blood, coincidences of platelets can be neglected [4]. Platelet counts measured with a haematology analyzer are added as a reference biomarker to highlight the kinetics of activated platelets.

Patient 1 suffered from a COVID-19 infection for at least 3 days when admitted to the ICU, and intubation was required (Fig. 4a). A secondary infection developed before ICU day 7, which compared well with the elevated levels of aggregates and an increase in the aggregate sizes. By day 8, the patient recovered from the viral infection, and the intubation could be completed with the PP aggregate levels and size distribution returning towards a healthy baseline. Compared to the viral pneumonia, patient 2 had a severe inflammation from an unknown infection and was intubated for over 5 days of blood sample measurements (Fig. 4b). Throughout the acute phase, the intubated patient showed stable aggregate levels around 5% and no large platelet size fraction. Patient 3 had a bacterial infection with intubation by day 3. Although the aggregate concentrations were not significantly elevated. Still, aggregate sizes were increased considerably (Fig. 4c). The PP aggregate concentration and size distributions results indicate that in-depth analysis potentially provides complementary early biomarker information on risks and correlates well with SOFA scores. Pneumonia patient 4, who was not intubated, shows a significant deviation from the prior patient cases with low SOFA scores despite elevated aggregate concentrations (Fig. 4d). Note that the occurrence of microaggregates having three or more platelets was high in the first 2 days of the ICU stay. The patient responded well to the antiviral therapy and was discharged to the general ward by day 6. The low SOFA score never showed a risk for organ dysfunction, but could also not reflect the acute phase of the ICU patient on day 2. Overall, platelet concentrations did not provide relevant information, such as platelet consumption or some kinetics related to the acute phase of the patients.

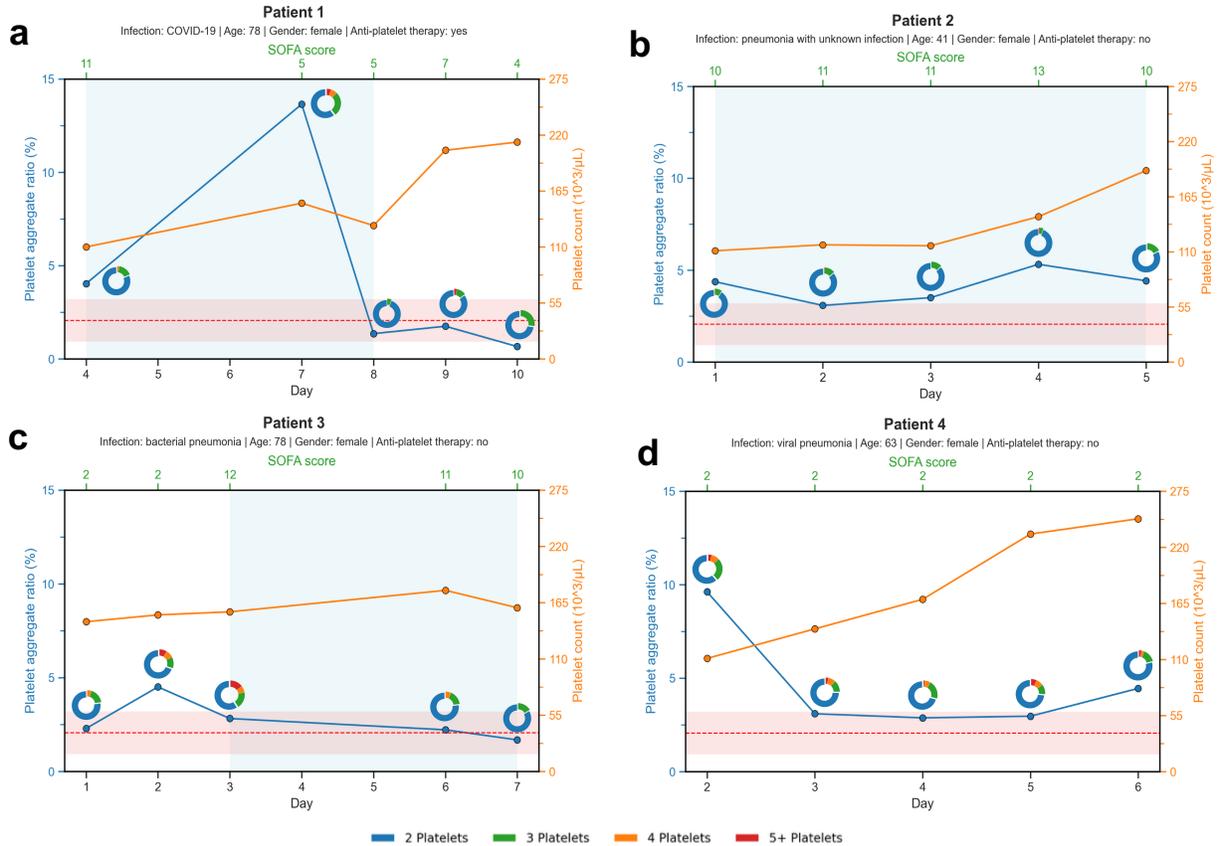

**Figure 4: Patient case study demonstrating RT-HAD's capability of quantifying platelet microaggregates as an early predictive biomarker for risk stratification.** Each plot (a-d) shows platelet aggregate ratio (blue), platelet counts (orange), and SOFA score across the days of patients' ICU stay. Pale blue regions indicate the intubation period, with the red band as the healthy reference for platelet microaggregates (mean ± SD). **(a)** Patient 1 - COVID-19: secondary infection drives a sharp rise in aggregate level and size that subsides once the viral infection resolves, anticipating extubation. **(b)** Patient 2 - pneumonia with unknown infection: aggregate concentration stays ~5% without large aggregates during prolonged intubation, mirroring persistent inflammation. **(c)** Patient 3 - bacterial pneumonia: surge in aggregate size, but not in concentration, precedes respiratory decline and intubation on day 3. **(d)** Patient 4 - viral pneumonia: early spikes in concentration and large aggregates highlight the acute phase despite low SOFA scores: values normalize with therapy before discharge to the general ward on day 6. Across all cases, microaggregate kinetics provide earlier and enhanced risk indicators than platelet counts or SOFA, supporting their use as prognostic biomarkers. The clinical patient information is detailed in Table S3.

## Discussion

We presented RT-HAD, an end-to-end real-time image and data processing deep learning framework developed specifically for DHM to realize label-free haematology analysis, enabling an enhanced biomarker panel by integrating holographic reconstruction, blood cell detection, and aggregate analysis. This pipeline allows high-throughput blood cell analysis with minimal computational hardware requirements. It eliminates the need for raw data storage, reducing storage size by over 99%, while addressing the long-standing "big data" problem in imaging flow cytometry for medical applications of DHM. OAH-Net [39] lies at the core of RT-HAD, a deep learning-based holographic reconstruction and phase retrieval model, that reconstructs over 10,000 holograms (>30 GB) per patient sample in under 5 ms per frame, that would otherwise require over an hour-long reconstruction process with traditional algorithms such as angular spectrum method (ASM) or iterative phase retrieval [46]. Given the various deep learning-based reconstruction models introduced in the [47-50] literature, these models often suffer from generalizing over samples out of their training set. Conversely, OAH-Net generalizes well across diverse phase targets and blood cells with structural similarity index (SSIM) scores exceeding 90%, significantly improving both the quality and speed of reconstructing phase images while maintaining inference time much lower than DHM camera's acquisition rate of 105 FPS [39].

For blood cell detection, RT-HAD processes full-frame phase images directly. This architectural design yields over 25% mAP50 improvement and >10-fold/frame speed-up in inference compared to processing raw holograms, which essentially fails to recognize small structures like platelets. We optimized inference by quantizing the YOLOv8x-p2 model into FP16, reducing memory usage and enabling processing with larger batch sizes. Despite the deeper architecture, YOLOv8x-p2 was chosen due to its superior performance in platelet detection while still reaching inference well under 10 ms combined with holographic reconstruction. Therefore, compared to prior art [4], RT-HAD speeds up image processing by >100-fold and dramatically increases the throughput of analysing more blood cells per inference run by a factor of >25-fold.

The graph-based aggregate analyzer identifies and quantifies blood-cell aggregates by constructing a spatial graph from detected cells as nodes. This module is model-agnostic and supports plug-and-play integration with any object-detection model. Consequently, our graph-based aggregate analysis module performed well in detecting true platelet aggregates, matching human experts' accuracy with only a marginal (<10%) deviation in clinical samples. In terms of analytical validity, Clinical Laboratory Improvement Amendments (CLIA) define the Total Allowable Error (TEa): the maximum permissible combined error used for method validation and accreditation in hematology as ±15% for leukocytes and ±25% for platelets [51]. Hence, results generated by RT-HAD fall within TEa bounds considered clinically acceptable for CBC parameters. Furthermore, conventional hematology analyzers raise flags when platelet clumps distort signals, prompting manual smear review and interrupting automation, yet they often miss true clumps [52]. In contrast, RT-HAD detects those rare microaggregates and treats them as functional biomarkers to enhance traditional hematology panels without workflow interruption. We finally demonstrated that longitudinal observation of biomarker kinetics from RT-HAD in single pneumonia patient cases indicates the potential to support clinical decision-making, using platelet microaggregates as a new set of predictive biomarkers in acute care. RT-HAD thus enables real-time, quantitative, label- and sample-preparation-free blood-cell analysis with an enriched biomarker panel at the point-of-care.

RT-HAD matches with a TAT of <1.5 min POCT requirements in acute care by eliminating raw data storage needs and enabling direct decision-making from processed phase images. It also complies with haematology regulatory standards where raw data archival is not mandatory, simplifying the deployment and reducing the carbon footprint of DHM imaging systems. Future improvements in CUDA and deep learning algorithms could lower computational load, making the system widely adopted even on average computer hardware, ideal for resource-limited settings. In summary, RT-HAD represents a significant architectural advance in real-time haematology imaging, which overcomes key limitations of existing imaging flow cytometers and delivers a scalable, cost-effective, and sustainable diagnostic platform to further accelerate DHM's clinical translation.

## Materials and methods

### DHM-based imaging flow cytometer

A customized digital holographic microscopy (Ovizio Imaging Systems, Belgium) was utilized to acquire the raw holographic frames. The optical setup employs a superluminescent light-emitting diode (SLED, OSRAM) with a centre wavelength of 528 nm that provides partially coherent Koehler illumination in transmission mode. A condenser lens assembly shapes the illumination to ensure uniform lighting across the field-of-view of the microfluidic channel, while the transmitted light interacts with the flowing blood cells. A Nikon CFI LWD 40× objective with numerical aperture (NA) of 0.55 was used to collect the scattered light from the sample and relay it onto a high-speed CMOS-based imager (Grasshopper GS3-U3-32S4, FLIR) with an acquisition speed of 105 frames/s. A custom microfluidic chip is integrated into the microscope to achieve high-throughput imaging. The chip made of Poly (methyl methacrylate) (PMMA) features a straight channel with cross-sectional dimensions of 50 μm × 100 μm that is designed to minimize shear stress, comparable with human veins not to disrupt fragile blood cells aggregates while ensuring that cells are viscoelastically focused in the channel to form a monolayer to eliminate the need for active autofocusing during the measurements.

### Data acquisition and dataset curation

To generate the training dataset, following obtaining written informed consent, venous whole blood from five healthy adult volunteers was drawn into EDTA blood tubes (BD Vacutainer). The different blood cells were isolated from whole blood. Erythrocytes and platelet-rich plasma (PRP) were isolated by gradient density centrifugation with high purity. In contrast, subtypes of leukocytes were isolated through magnetic-activated cell sorting (MACS) to obtain highly pure cell populations (Miltenyi Biotec MACSxpress®). Neutrophils, lymphocytes (B and T cells), and eosinophils were isolated directly from whole blood with one step. However, monocytes were isolated by obtaining peripheral blood mononuclear cells (PBMC), followed by selective isolation using MACS.

Furthermore, chemically induced platelet aggregates were generated to validate and optimize algorithms for detecting and characterizing these aggregates. Platelets were isolated with high purity and viability, followed by a 15-min incubation with thrombin receptor-activating peptide (TRAP). Mechanical stress, simulating physiological conditions, was applied by vortexing the platelet suspension at 3,000 rpm for 1 min. This induced rapid platelet aggregation of varying sizes and shapes. Similarly, leukocyte-platelet aggregates were created in vitro to study intercellular interactions and aggregate formation. Based on the method for generating synthetic platelet aggregates, the protocol was modified to include leukocytes, specifically neutrophils or monocytes [16]. TRAP-activated platelets to stimulate aggregation and enhance their adhesion to leukocytes. The cell suspensions were mixed and subjected to mechanical stress by vortexing at maximum speed for 1 min. This agitation resulted in leukocyte and platelet aggregates of various shapes and sizes, such as NETs and pseudopodia formations in neutrophils and platelets.

Raw holographic data from different experiments of isolated blood cell types were initially reconstructed into phase and amplitude images for annotation using OAH-Net. For the isolated or stimulated blood cells, at least two measurements (≥10,000 frames each) per donor were recorded, and the frames without cells were discarded before annotation. A custom tool was developed to label these images by drawing bounding boxes around individual cells and assigning main classes (i.e., erythrocytes, leukocytes, platelets) curated by annotators who had domain knowledge in blood cell analysis and were trained by a medical doctor from the anesthesiology department. We followed a verification workflow where an annotator performed the initial annotation, and two additional annotators independently reviewed. Inter-annotator reliability was evaluated with the two-way random-effects, absolute-agreement, single-measure intraclass correlation coefficient - ICC (2, 1) [53]. Pooled across all images and cell types, ICC was 0.99 (95% CI 0.99 - 0.99, F=340.8, $p<0.001$), indicating excellent concordance among the three annotators. When computed separately by cell class, agreement remained excellent for erythrocytes (ICC=0.99), suitable for platelets (ICC=0.81), and moderate for leukocytes (ICC=0.75) according to the Koo-and-Li interpretation thresholds [54]. The relatively lower leukocyte and platelet ICCs likely reflect the interpretation differences of the broader morphological variations associated with various cellular pathological states (e.g., activation) among annotators. Nevertheless, all F-tests were highly significant ($p<0.001$), confirming that the observed agreement among annotators was well established. Therefore, discrepancies among annotators during the annotation process were resolved through discussion on comparable cell classes and morphologies until consensus was reached.

The preliminary dataset of modest size (e.g., 500 frames) was used to train the object detection model. Once the model achieved satisfactory accuracy, it was deployed to annotate additional images. The annotations produced by the model were then manually reviewed and corrected. This iterative process continued until a comprehensive dataset of over 95,000 frames (training set: 76,063, validation set: 19,016, holdout test set: 1,512), covering erythrocytes, leukocytes (e.g., neutrophils, monocytes, lymphocytes (B and T cells), and eosinophils), synthetic aggregates, as well as clinical patient samples where the cell classes indicated as aggregates contain images collected from stimulation experiments or patient sample measurements), all other cell classes contain single cells of different kinds. Therefore, our annotations are based on preliminary model predictions followed by human corrections. We performed patient-level data splitting into training, validation, and testing to prepare the dataset for machine learning. All our results reported in this paper are based on holdout testing: *i)* training set: control (n=10), pneumonia (n=10) and fever (n=10) patients, *ii)* validation set: control (n=10), pneumonia (n=10) and fever (n=10) patients and *iii)* holdout testing set: control (n=10), pneumonia (n=5) and fever (n=18). Table S1 details the number of frames for each cell type and data splitting.

The curated dataset was then utilized for training and optimizing the final object detection model. this manuscript reported the results on the holdout test set.

The blood samples from acute care (n=110) and day surgery patients (n=20) from the ICU were drawn into blood tubes containing citrate anticoagulants (BD Vacutainer). The blood samples were diluted (1:100) in Phosphate Buffered Saline (PBS) with an additional 0.05% (w/v) of Polyethylene Oxide (PEO, with molecular weight of 4,000,000, Sigma Aldrich) solution to eliminate cell overlaps and facilitate the alignment of cells along the channel centreline. Acquisition of 10,000 holographic frames typically takes ~90 s at 105 frames per second per measurement for each patient. All blood samples used in this work were deidentified and obtained through adherence to the Domain Specific Institutional Review Board with numbers 2021-00930 (approval date: 17/12/2021) and 2021-01130 (approval date: 17/2/2022) from National University Hospital, Singapore.

**Architecture and implementation of RT-HAD**
RT-HAD is an end-to-end deep learning framework that seamlessly integrates three main modules: *i)* hologram reconstruction and phase retrieval, *ii)* single blood cell detection, and *iii)* identification of blood cell aggregates. OAH-Net [39] relies on two separate neural networks specializing in the spatial filtering of object waves in $x$ and $y$ components of holograms (shear interferometer-based DMH used in this work), rescaling and separating complex-valued data into phase and amplitude images. The architecture is a physics-consistent neural network to eliminate hallucination and leverage external generalization over sample types not seen during training by the model. This network employs a Fourier Imager Head (FIH) to perform a learnable filtration in the Fourier spectrum. In the frequency domain, the off-axis interference term is isolated using trainable two-dimensional filter matrices that act similarly to a circular band-pass filter. The network then applies an inverse Fourier transform to yield the complex optical field, from which both amplitude and phase images are computed. A subsequent phase unwrapping layer removes $2\pi$ ambiguities, ensuring a continuous phase map. The training loss for this module is the difference in amplitude and phase modalities, and a perceptual loss to ensure that the reconstructed images match high-level structural features of the ground truth.

The second module is a hybrid graph-based convolutional neural network. YOLOv8x-p2 processes the reconstructed phase images to identify individual blood cells (e.g., erythrocytes, leukocytes, platelets). This variant is chosen because it builds on the robust YOLOv8x backbone, which utilizes advanced Cross-Stage Partial (CSP)-based feature extraction with Cross-Stage Partial with Two-Fusion (C2f) modules. These modules capture fine-grained details essential for recognizing the subtle differences in cell morphology. YOLOv8x-p2 integrates an extra P2 detection layer: a high-resolution output branch that operates on feature maps with a lower downsampling factor (e.g., stride four instead of the conventional stride 8). This higher spatial resolution preserves minute details often lost in deeper layers, significantly improving the detection performance on small objects such as single platelets or those embedded within aggregates. In addition, the model employs a decoupled detection head that independently optimizes bounding box regression and classification. This design enhances localization precision and ensures that even the smallest objects are correctly identified.

Furthermore, our blood cell detection and aggregate analysis module benefits from a hybrid design for post-processing by identifying aggregates following object detection. A spatial graph for each frame is generated where nodes represent each detected cell by YOLOv8x-p2 and edges connect neighbouring cells within a pre-defined proximity threshold. A pairwise distance matrix for target cell groups (i.e., leukocytes and platelets) is calculated based on each adjacent cell's centroid distances. This approach distinguishes between single cells and those in aggregates and categorizes aggregates by type (e.g., PP, LP, LL) and returns the number of constituent cells.

To further reduce latency, we implemented batch processing (with a batch size of 130) using PyTorch's JIT compiler and NVIDIA TensorRT, which enables quantization of model weights. We deployed quantization of complete precision (FP32) model weights to half precision (FP16) to reach faster inference. Quantization is unavailable since the aggregate analyzer module relies on graph construction and arithmetic operations. Using GPU acceleration, our framework was implemented in Python (v3.12) with PyTorch (v2.5.1) library. Each model within the framework was integrated to minimize data movement and latency for high efficiency and low inference time for real-time processing. The experiments were performed on a desktop machine equipped with graphics processing units (GPUs) of 2 x Asus RTX4090 24 GB, the central processing unit (CPU) of Ryzen Threadripper Pro 5965 WX with 128 GB of RAM.

**Training of the deep learning framework**

OAH-Net [39] was trained end-to-end in a supervised manner using blood cell samples. Input holograms were recorded with and without the sample. The target images (ground truth) were phase ($\phi$) and amplitude ($A$) images reconstructed using the Fourier transformation and spatial filtering techniques. The target images were generated using the microscope manufacturer's software. Autofocusing was not implemented, as the viscoelastic focusing utilized in imaging the microfluidic chip at the DHM set ensures >95% of cells were correctly focused. Due to data imbalance between background and sample pixels, a weighted L1 loss function was used to emphasize areas with higher importance. OAH-Net was trained using the Adam optimizer with a constant learning rate optimized by grid search, and training stopped after 200 epochs without improvement in validation loss. The model with the lowest validation loss was selected for testing.

We trained the YOLOv8x-p2 architecture from scratch on our custom dataset for the object detection network. During the training, the dataset was randomly split 80%-20% into training and validation sets by frames, and we used the default hyperparameters provided in the Ultralytics library (v8.3.91) without hyperparameter tuning. We deployed a learning rate of 0.01 and a weight decay term of 0.0005 for each gradient step with an L2 regularization term to help prevent overfitting. A batch size of 64 was chosen to balance efficient computation with stable gradient updates, and the training was performed for 10 epochs. An extensive data augmentation strategy was applied to improve the model's performance. Sample images from our dataset were augmented using horizontal and vertical flips and random rotations sampled from a uniform distribution between +90 and -90 degrees. These augmentations helped increase the training data's diversity, thereby enhancing the model's robustness and ability to generalize to specific small single platelets and platelet aggregate structures. Throughout the training process, the model was continuously monitored on the validation set, and the best-performing model, defined by the lowest validation loss, was saved as the final version. The saved model's performance was further validated on a separate holdout test set generated through a random five-fold split of 1,512 frames drawn from 33 patient samples in the acute care and healthy cohorts that had not been used in the training and validation set (Table S1).

## Data availability

A representative dataset, consisting of raw, containerized measurements from a healthy adult volunteer, holograms, corresponding reconstructed phase and amplitude images, and annotation files, is available at https://zenodo.org/records/15338907. Patient data acquired retrospectively under institutional approval remains subject to access restrictions and cannot be made publicly accessible due to data privacy considerations. Researchers seeking access to raw or processed patient data should direct their requests to the corresponding authors; such data may be shared upon reasonable request.

## Code availability

We detail every deep-learning method and software library employed in this study, while keeping the paper accessible to clinicians and non-specialist scientists. The custom-developed code used in this work is available at https://github.com/CellFace/rt_had.

## Acknowledgements

This research is supported by the National Research Foundation, Prime Minister's Office, Singapore, under Intra-CREATE Thematic Grant (NRF2019-THE002-0008). The figures were partly generated using bioRender (www.biorender.com).

## Author contributions

M.E.C. and O.H. conceived and designed the study. W.S.K., J.T.Y.S., and M.E.C. coordinated patient recruitment and sample collection. J.K. and M.S. developed cell isolation protocols. K.De. performed DHM measurements, experimental data collection, and curation of the training dataset. K.De., Q.C., L.W., and S.K.M. developed the image and data processing framework. K.De. and Q.C. performed the data analysis. K.De. and O.H. designed the figures and tables. G.S., C.P.d.C., P.A.K., L.R., M.E.C., H.K.L., K.Di., and O.H. supervised the study. K.De. drafted the initial manuscript. All authors contributed intellectual content while drafting and revising the work, reviewing and approving the final version.

## Competing interests

K.D., Q.C., L.W., S.K.M., H.K.L., and O.H. are inventors of a pending patent related to this work filed by the Singapore Patent Office (10202401892U). The authors declare that they have no other competing interests.

## Additional information

Correspondence and material requests should be addressed to Kerem Delikoyun and Oliver Hayden.

## Supplementary figures & tables:

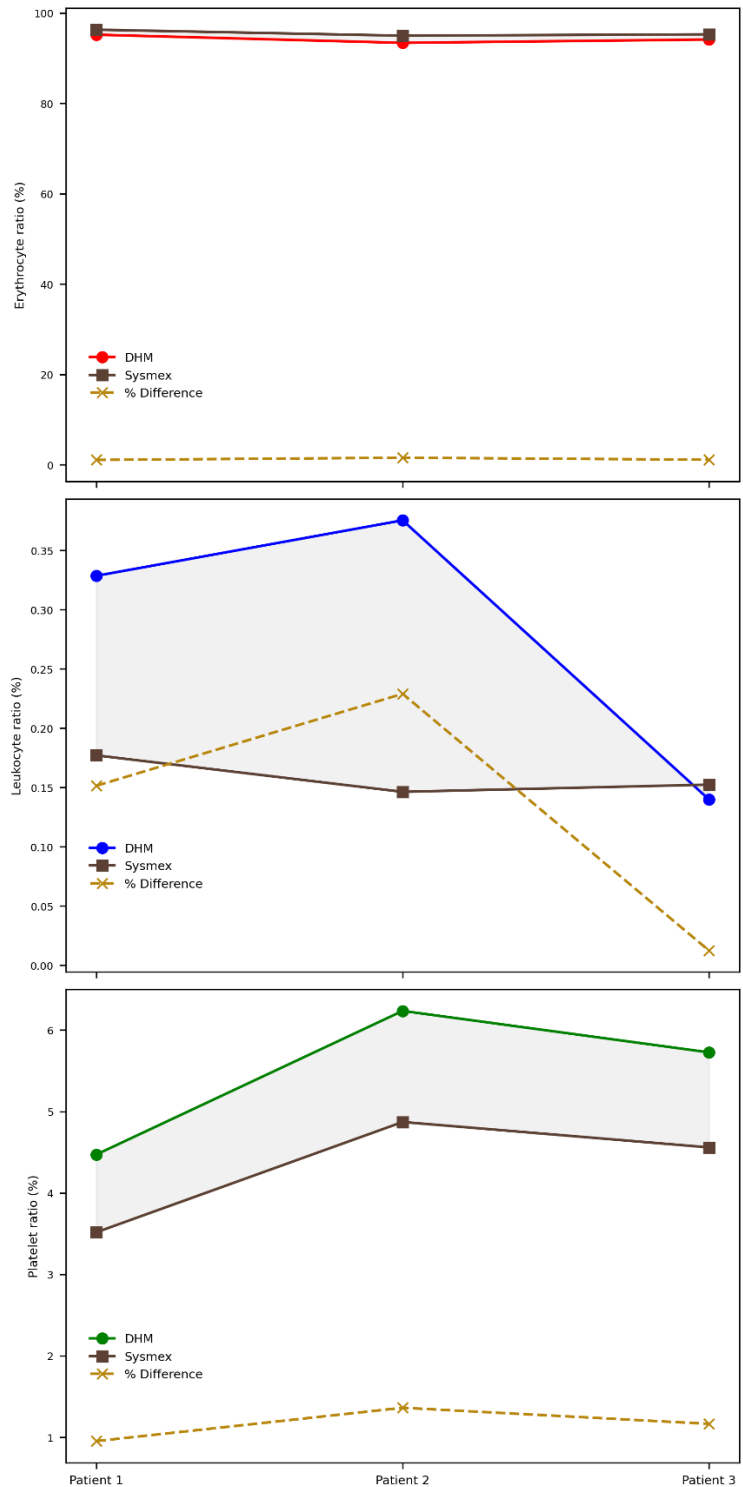

**Figure S1:** The comparison of blood cell ratios in whole blood between RT-HAD and an automated haematology analyzer (Sysmex XN-350, Japan) from the holdout test set, highlighting differences of less than 1% in erythrocytes and platelets, and less than 0.2% in leukocytes.

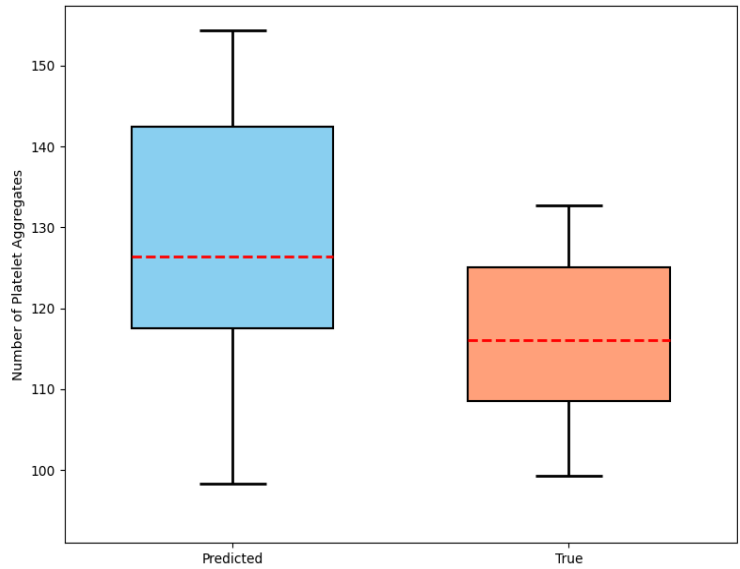

**Figure S2:** Analysis of clinical patient control sample for platelet aggregates by RT-HAD against human annotators, indicating prediction range with a mean error rate of 8.9% with three repeated measurements on the clinical holdout test set, red dashed lines indicate the mean.

**Table S1 | The composition of the dataset, including annotated full-size phase images from various types of leukocytes, synthetically generated aggregates and clinical samples.**

| Training set | |
|---|---|
| **Cell type** | **Number of frames** |
| B lymphocytes | 8,166 |
| Clinical samples | 8,430 |
| Clinical samples (aggregates) | 947 |
| Eosinophil | 2,378 |
| Erythrocytes | 8,092 |
| Monocyte | 4,348 |
| Neutrophil | 12,188 |
| Neutrophil (aggregates) | 1,553 |
| Platelet | 19,625 |
| Platelet (aggregates) | 2,175 |
| T lymphocytes | 8,161 |
| **Total** | **76,063** |
| Validation set | |
| **Cell type** | **Number of frames** |
| B lymphocytes | 1,999 |
| Clinical samples | 2,070 |
| Clinical samples (aggregates) | 236 |
| Eosinophil | 594 |
| Erythrocytes | 2,016 |
| Monocyte | 1,065 |
| Neutrophil | 3,183 |
| Neutrophil (aggregates) | 408 |
| Platelet | 4,883 |
| Platelet (aggregates) | 524 |
| T lymphocytes | 2,038 |
| **Total** | **19,016** |
| Holdout test set | |
| **Cell type** | **Number of frames** |
| Clinical samples | 1,412 |
| Monocyte (aggregates) | 30 |
| Platelet (aggregates) | 70 |
| **Total** | **1,512** |
| Distribution of patient numbers in the dataset | |
| **Patient Cohort** | **Train/valid/test** |
| Control | 10/10/10 |
| Pneumonia | 10/10/5 |
| Fever | 10/10/18 |

**Table S2 | Model performance and inference times across different models in YOLO family on the holdout test set comprised of full frame phase images from single erythrocytes, leukocytes, platelets, and their blood cell aggregates.**

| Model | Params (M) | Precision-all (%) | Precision-PLT (%) | mAP50-all (%) | mAP50-PLT (%) | Speed (ms) |
|---|---|---|---|---|---|---|
| YOLOv8n | 3.2 | 87.7 ± 2.6 | 94.1 ± 2.8 | 94.2 ± 1.6 | 93.8 ± 1.3 | 2.0 |
| YOLOv8s | 11.2 | 88.9 ± 4.0 | 93.7 ± 2.7 | 95.4 ± 1.9 | 95.4 ± 0.6 | 2.0 |
| YOLOv8m | 25.9 | 91.7 ± 2.7 | 95.1 ± 1.4 | 96.4 ± 1.1 | 95.4 ± 1.2 | 2.6 |
| YOLOv8l | 43.7 | 91.6 ± 3.9 | 95.5 ± 2.3 | 96.5 ± 1.0 | 96.0 ± 0.8 | 3.4 |
| YOLOv8x | 68.2 | 93.2 ± 2.2 | 96.0 ± 1.6 | 96.1 ± 1.1 | 95.9 ± 1.0 | 4.7 |
| *YOLOv8x-p2* | 66.6 | 93.8 ± 1.4 | *96.8 ± 2.6* | *96.3 ± 0.6* | *96.8 ± 0.8* | 6.6 |
| YOLOv9t | *2.0* | 87.2 ± 3.5 | 93.5 ± 3.6 | 94.3 ± 2.0 | 94.4 ± 1.5 | 2.0 |
| YOLOv9s | 7.2 | 93.4 ± 2.7 | 95.0 ± 1.8 | 96.1 ± 0.9 | 94.9 ± 0.8 | 2.0 |
| YOLOv9m | 20.1 | 91.8 ± 1.5 | 95.6 ± 2.8 | 95.8 ± 1.2 | 95.9 ± 0.6 | 3.2 |
| YOLOv9c | 25.5 | *94.1 ± 2.2* | 95.5 ± 2.1 | 96.2 ± 1.1 | 95.4 ± 1.4 | 3.2 |
| YOLOv9e | 58.1 | 89.7 ± 3.3 | 93.0 ± 2.4 | 95.8 ± 1.2 | 95.2 ± 0.7 | 6.3 |
| YOLOv10n | 2.3 | 91.7 ± 2.5 | 89.5 ± 4.4 | 94.1 ± 1.8 | 93.0 ± 1.9 | *1.6* |
| YOLOv10s | 7.2 | 92.1 ± 3.2 | 88.1 ± 6.1 | 95.5 ± 1.1 | 95.0 ± 0.8 | 1.7 |
| YOLOv10m | 15.4 | 91.6 ± 2.4 | 89.2 ± 4.1 | 95.7 ± 0.6 | 94.5 ± 1.0 | 2.3 |
| YOLOv10l | 24.4 | 94.0 ± 1.4 | 91.9 ± 5.1 | 96.1 ± 1.0 | 95.8 ± 0.9 | 3.2 |
| YOLOv10x | 29.5 | 92.0 ± 1.7 | 91.6 ± 3.9 | 95.8 ± 1.7 | 95.9 ± 1.3 | 4.3 |
| YOLOv11n | 2.6 | 90.1 ± 3.2 | 88.3 ± 5.2 | 94.2 ± 1.8 | 92.4 ± 1.5 | 2.0 |
| YOLOv11s | 9.4 | 91.3 ± 2.7 | 91.1 ± 3.8 | 95.3 ± 1.0 | 94.0 ± 1.4 | 2.0 |
| YOLOv11m | 20.1 | 92.0 ± 1.8 | 90.7 ± 1.8 | 96.3 ± 0.6 | 95.5 ± 0.7 | 2.7 |
| YOLOv11l | 25.3 | 93.0 ± 2.1 | 93.3 ± 3.6 | 96.1 ± 0.9 | 95.4 ± 1.1 | 3.0 |
| YOLOv11x | 56.9 | 93.6 ± 2.0 | 91.2 ± 4.7 | 96.0 ± 1.2 | 95.2 ± 1.2 | 4.5 |
| RT-DETR-L | 4.8 | 86.5 ± 4.2 | 81.4 ± 2.4 | 86.8 ± 1.4 | 83.9 ± 3.2 | 3.9 |
| RT-DETR-X | 6.9 | 86.8 ± 2.4 | 84.8 ± 7.0 | 88.8 ± 3.4 | 89.3 ± 1.6 | 5.3 |

**Table S3 | The descriptive clinical information of pneumonia patients from the intensive care unit presented in Figure 4.**

|  | Patient 1 | Patient 2 | Patient 3 | Patient 4 |
|---|---|---|---|---|
| **Demographics** | | | | |
| Age | 78 | 41 | 72 | 63 |
| Gender | Female | Female | Male | Female |
| **Comorbidities** | | | | |
| Diabetes mellitus (DM) | Uncomplicated | No | No | No |
| Cerebrovascular accident (CVA) | No | No | No | No |
| Myocardial infarction (MI) | No | No | No | No |
| **Therapy** | | | | |
| Antiplatelet | Yes | No | No | No |
| **Outcome** | | | | |
| Infection type | Viral | Unknown | Bacterial | Viral |
| Disposition | Ward | ICU | Ward | Ward |
| Hospital length of stay (day) | 17 | 17 | 126 | 11 |
| Admission day to ICU | 4 | 1 | 1 | 2 |
| Admission day to ward | N/A | N/A | N/A | 6 |
| Intubation duration (day) | 5 | 10 | 27 | 0 |